\magnification\magstephalf
\overfullrule 0pt
\def\D{\Delta}

\font\rfont=cmr10 at 10 true pt
\def\ref#1{$^{\hbox{\rfont {[#1]}}}$}


\font\fourteenbf=cmbx12 scaled\magstep1

\font\tenbfit=cmbxti10
\font\sevenbfit=cmbxti10 at 7pt
\font\fivebfit=cmbxti10 at 5pt
\newfam\bfitfam 
\textfont\bfitfam=\tenbfit  \scriptfont\bfitfam=\sevenbfit
\scriptscriptfont\bfitfam=\fivebfit

\font\eightit=cmti8

\font\tenbfit=cmbxti10
\font\sevenbfit=cmbxti10 at 7pt
\font\fivebfit=cmbxti10 at 5pt
\newfam\bfitfam 
\textfont\bfitfam=\tenbfit  \scriptfont\bfitfam=\sevenbfit
\scriptscriptfont\bfitfam=\fivebfit

\font\tenbit=cmmib10
\newfam\bitfam
\textfont\bitfam=\tenbit%

\font\tenmbf=cmbx10
\font\sevenmbf=cmbx7
\font\fivembf=cmbx5
\newfam\mbffam
\textfont\mbffam=\tenmbf \scriptfont\mbffam=\sevenmbf
\scriptscriptfont\mbffam=\fivembf

\font\tenbsy=cmbsy10
\newfam\bsyfam 
\textfont\bsyfam=\tenbsy%

  \def\g {\gamma} 
\def\e{\epsilon}

\def\pd {\partial}
\def\pmb#1{\setbox0=\hbox{#1}
 \kern.05em\copy0\kern-\wd0 \kern-.025em\raise.0433em\box0 }

\def\slash{/\kern-.5em}

\def \half {{\textstyle {1 \over 2}}}

\def \quarter {{\textstyle {1 \over 4}}}

 %


\def\boxit#1{\vbox{\hrule\hbox{\vrule\kern1pt\vbox
{\kern1pt#1\kern1pt}\kern1pt\vrule}\hrule}}

\def\h{\hfill\break}
\parskip=6pt
\parindent=0pt
\hsize=17truecm\hoffset=-5truemm
\vsize=22truecm
\def\footnoterule{\kern-3pt
\hrule width 17truecm \kern 2.6pt}


\catcode`\@=11 

\def\nolabels{\def\wrlabeL##1{}\def\eqlabeL##1{}\def\reflabeL##1{}}
\def\writelabels{\def\wrlabeL##1{\leavevmode\vadjust{\rlap{\smash%
{\line{{\escapechar=` \hfill\rlap{\sevenrm\hskip.03in\string##1}}}}}}}%
\def\eqlabeL##1{{\escapechar-1\rlap{\sevenrm\hskip.05in\string##1}}}%
\def\reflabeL##1{\noexpand\llap{\noexpand\sevenrm\string\string\string##1}}}
\nolabels
\global\newcount\refno \global\refno=1
\newwrite\rfile
\def\defref{$^{{\hbox{\rfont [\the\refno]}}}$\nref}
\def\nref#1{\xdef#1{\the\refno}\writedef{#1\leftbracket#1}%
\ifnum\refno=1\immediate\openout\rfile=refs.tmp\fi
\global\advance\refno by1\chardef\wfile=\rfile\immediate
\write\rfile{\noexpand\item{#1\ }\reflabeL{#1\hskip.31in}\pctsign}\findarg}
\def\findarg#1#{\begingroup\obeylines\newlinechar=`\^^M\pass@rg}
{\obeylines\gdef\pass@rg#1{\writ@line\relax #1^^M\hbox{}^^M}%
\gdef\writ@line#1^^M{\expandafter\toks0\expandafter{\striprel@x #1}%
\edef\next{\the\toks0}\ifx\next\em@rk\let\next=\endgroup\else\ifx\next\empty%
\else\immediate\write\wfile{\the\toks0}\fi\let\next=\writ@line\fi\next\relax}}
\def\striprel@x#1{} \def\em@rk{\hbox{}} 
\def\lref{\begingroup\obeylines\lr@f}
\def\lr@f#1#2{\gdef#1{\defref#1{#2}}\endgroup\unskip}
\newwrite\lfile
{\escapechar-1\xdef\pctsign{\string\%}\xdef\leftbracket{\string\{}
\xdef\rightbracket{\string\}}}

\def\writestop{\def\writestoppt{\immediate\write\lfile{\string\p
ageno%
\the\pageno\string\startrefs\leftbracket\the\refno\rightbracket%
\string\def\string\secsym\leftbracket\secsym\rightbracket%
\string\secno\the\secno\string\meqno\the\meqno}\immediate\closeout\lfile}}
\def\writestoppt{}\def\writedef#1{}
\catcode`\@=12 
\rightline{DAMTP-1999-66 \ \ \ \ TUW 99/10}
\vskip 6pt
\centerline{\fourteenbf QCD PRESSURE AND THE TRACE ANOMALY}
\vskip 6mm
\centerline{\bf I T Drummond, R R Horgan, P V Landshoff}
\vskip 1mm
\centerline{DAMTP, University of Cambridge$^\star$}
\vskip 3mm
\centerline{\it and}
\vskip 3mm
\centerline{\bf A Rebhan}
\vskip 1mm
\centerline{Institut f\"ur Theoretische Physik, Technische Universit\"at Wien,
Vienna$^\star$}
\footnote{}{$^\star$ itd@damtp.cam.ac.uk \  rrh@damtp.cam.ac.uk \
pvl@damtp.cam.ac.uk \ rebhana@hep.itp.tuwien.ac.at}
\vskip 0.5 cm
{\bf Abstract}

Exact relations between the QCD thermal pressure and the trace anomaly
are derived. These are used, first, to prove the equivalence of the
thermodynamic and the hydrodynamic pressure in equilibrium in the
presence of the trace anomaly, closing a gap in previous arguments.
Second, in the temporal axial gauge a formula is derived which expresses
the thermal pressure in terms of a Dyson-resummed two-point function.
This overcomes the infrared problems encountered in the conventional
perturbation-theory approach.

\vskip 5mm

{\bf 1 Introduction}

Conventional methods of calculating the QCD thermal pressure in perturbation
theory have encountered an apparently insuperable infrared 
problem\defref\braaten{
E Braaten and A Nieto, Physical Review Letters 76 (1996) 1417
and Physical Review D53 (1996) 3421 and references therein
}. 
In previous papers, we have suggested a resummation technique that
may 
overcome this problem and tested it on scalar field theory\defref\foam{
I T Drummond, R R Horgan, P V Landshoff and A Rebhan, Physics Letters 
B398 (1997) 326 and Nuclear Physics B524 (1998) 579\h
D B\"odeker, O Nachtmann, P V Landshoff and A Rebhan, 
Nuclear Physics B539 (1999) 233}.
Although in principle this technique can be applied to QCD, in practice this
would be far from simple, since it involves introducing a mass for the gluon.
Although this mass must disappear at the end of the calculation, its
presence at intermediate stages is a considerable inconvenience. In this paper,
therefore, we introduce a more suitable technique for performing the
resummation that is needed to overcome the infrared problem.

It is common to distinguish two different definitions of 
pressure\defref\jizba{
P Jizba, hep-th/9801197
}.
The first, called the thermodynamic pressure, is given in terms of the
grand partition function 
$$
Z=\sum _i\langle i|\exp (-\beta H)|i\rangle
\eqno(1.1)
$$
The summation is over a complete set of physical states $i$.
The Hamiltonian $H$ is an integral over an element of the energy-momentum
tensor:
$$
H=\int d^{n-1}{\bf x}\; T^{00}(x)
\eqno(1.2)
$$
The thermodynamic pressure is defined to be
$$
P=TV^{-1}\log Z
\eqno(1.3)
$$

The other definition of the pressure is in terms of 
the thermal average of another element of the energy-momentum
tensor. The hydrodynamic pressure is
$$
{\cal P}=\langle T^{33}\rangle
\eqno(1.4)
$$
Although the two pressures $P$ and $\cal P$ are generally expected
to be equal for a system in thermal equilibrium\ref{\jizba},
it was remarked some time ago by Landsman and van Weert\defref\landsman{
N P Landsman and C van Weert, Physics Reports 145 (1987) 141
}
that the presence of the trace anomaly might cause certain complications.

In section 2 we derive formulae for the derivatives of the thermodynamic
pressure $P$ with respect to the bare coupling $g$ and to the temperature:
see (2.7) and (2.8). 
In section 3 we replace $g$ with the renormalised coupling $g_R$,
and we use the renormalisation group to show that the two formulae
for the derivatives of the thermodynamic pressure $P$ together imply that 
$P$ is equal to the hydrodynamic pressure $\cal P$.
We show also that the pressure's departure from simple $T^4$ behaviour
is calculable from the QCD trace anomaly. 
In section 4 we show, by working in the temporal gauge, 
how simple Dyson resummation of the thermal gluon propagator removes
the infrared problem. In section 5 we derive some simple results
for the small-coupling limit, which will be the starting point for
a perturbation expansion. We plan to consider such an expansion in a further
paper. 
\bigskip
{\bf 2 Unrenormalised formulae}

Consider pure-glue QCD for simplicity.
At finite temperature there is a choice: one may
heat up only the physical components of the gauge field\defref\toni{
P V Landshoff and A Rebhan, Nuclear Physics B383 (1992) 607
},
but for our purposes it will be more convenient to adopt the more
conventional formalism where also the unphysical components are heated.
Then one may write\defref\lebellac{
M Le Bellac, {\sl Thermal field theory}, Cambridge University Press (1996)
} 
the grand partition function as a path integral:
$$
Z(g,T)=\int dA\, dc\, d\bar c\,dB\,\exp\Big [i\int ^{\tau-i\beta}_{\tau}dx^0
\int d^{n-1}{\bf x}\, {\cal L}(g,x)\Big ]
\eqno(2.1)
$$
The Lagrangian is
$$
{\cal L}(g,x)=-\quarter F^2 +{\cal L}_{\hbox{{\sevenrm GF}}}+
{\cal L}_{\hbox{{\sevenrm GHOST}}}
\eqno(2.2)
$$
We have chosen to implement the gauge fixing by means of an auxiliary field $B$.
The two common choices of $\tau$ are $\tau =0$, with the $x^0$ integration
running along the imaginary axis, which is the imaginary-time formalism,
and $\tau=-\infty$, with the $x^0$ integration following the Keldysh
contour, which is a version of the real-time formalism\ref{\lebellac}.
Either formalism may be
used for most of our work, though when a definite choice has to be made
we will choose the second. So far, the fields are unrenormalised. 

We scale each field by some power of $g$, so obtaining new fields
$^*\!A,\,^*c,\,^*\bar c,\,^*\!B$.
The integration measure then acquires a power
of $g$ that depends, loosely speaking,  on the number of space-time points,
so it is $T$-dependent. In order to cancel this, we consider 
the ratio
$$
{Z(T,g)\over Z(T,0)}
\eqno(2.3)
$$
and make the same field transformation in the denominator as in the numerator,
so that the extra factor cancels. We then differentiate 
the logarithm of (2.3) with respect to
$g$, and transform back to the original fields. Choosing different powers
of $g$ in the definitions of the starred fields and using the fact that the
derivative of (2.3) must be independent of what powers we choose, we may
obtain various identities. For our purposes, the most useful change
of field variables is
$$
A=^*\!\!A/g~~~~~~c=^*\!c~~~~~\bar c=^*\!\bar c~~~~~B=^*\!\!Bg
\eqno(2.4)
$$
We use the definition (1.3) of the thermodynamic the pressure 
in terms of the grand partition
function, and the translation invariance property  that
$$
\Big\langle \int ^{\tau-i\beta}_{\tau}dx^0
\int d^{n-1}{\bf x} \,F^2(x)\Big\rangle=-i\beta V \langle F^2(0)\rangle
\eqno(2.5)
$$
where $V$ is the volume of the $(n-1)$-dimensional ${\bf x}$-space.
We find that
\def\F{\hbox{{\fiverm FREE}}}
$$
\hat P(T,g)=P(T,g)-P_{\F}(T)
\eqno(2.6)
$$
satisfies 
$$
{\pd\over\pd g}\hat P(T,g)=
{1\over 2g}\Big [\langle F^2(0)\rangle -\langle F^2(0)\rangle_{\F}\Big ]
\eqno(2.7)
$$
Notice the importance of the subtraction term: it avoids a divergence
at $g=0$.

We may directly obtain 
a different derivative of $\hat P(T,g)$.
We use the original definition (1.1) of $Z$,
together with the
expression (1.2) for the Hamiltonian $H$ as an integral over the energy
density $T^{00}$:
$$
{\pd\over\pd\beta}\big [\beta \hat P(T,g)\big ]
=-\Big [\langle T^{00}(0)\rangle-\langle T^{00}(0)\rangle_{\F}\Big ]
\eqno(2.8)
$$
where again we have used translational invariance.
 
A gauge-invariant form for the gluonic part
of the energy-momentum tensor is
$$
T^{\mu\nu}=-F^{\mu\rho}F^{\nu}_{\;\rho}+\quarter g^{\mu\nu}F^2
\eqno(2.9)
$$
A trace over colour indices is understood. 
Classically, $T^{\mu\nu}$ is traceless, but in quantum theory there
is an anomaly that changes this.  We use dimensional regularisation
so that we work intially  $n=4-\e$ dimensions. Then
$$T^\mu_{\;\;\mu}=-\quarter \e F^2
\eqno(2.10)
$$
Renormalisation leaves behind a nontrivial limit as $\e\to0$.

Introduce the notation
$$
D^M=\langle F^{M\rho}F^M_{\;\;\;\rho}\rangle -
\langle F^{M\rho}F^M_{\;\;\;\rho}\rangle
_{\F}
\eqno(2.11)
$$
where $M$ is not summed. Then 
$$\eqalignno{
{\pd\over\pd g}\hat P(T,g) &= {1\over 2g}(D^0-(3-\e )D^3)&(2.12a)\cr
{\pd\over\pd\beta}\big [\beta \hat P(T,g)\big ]
 &={\textstyle{3\over 4}}D^0 +\quarter (3-\e )D^3
&(2.12b)\cr}
$$
in $n=4-\e$ dimensions.
\bigskip\goodbreak
{\bf 3 Renormalisation}

In (2.12), the coupling $g$ is unrenormalised, and both $D^0$ and $D^3$
are divergent. We now replace $g$ with a renormalised coupling $g_R$.
For this,
we use dimensional regularisation and the MS scheme,
so that we define the $g_R$ by
$$
g=\mu ^{\e /2}Z(g_R)g_R(\mu)
\eqno(3.1a)
$$
where $Z(g_R)$ is a combination of wave-function and vertex renormalisation
factors,
and a beta function by
$$
\beta (g_R)=\mu{\pd g_R\over\pd\mu}\Big\arrowvert _g
\eqno(3.1b)
$$
The beta function has the structure\defref\grinstein{
G 't Hooft, Nuclear Physics B61 (1973) 455\hfill\break
J C Collins, Nuclear Physics B80 (1974) 341\hfill\break
B Grinstein and L Randall, Physics Letters B217 (1989) 335
}
$$
\beta (g_R)=-\half\e g_R+\tilde\beta (g_R)
\eqno(3.2a)
$$
where for small $g_R$
$$
\tilde\beta (g_R)\sim \beta_0g_R^3
\eqno(3.2b)
$$
Differentiating (3.1a) with respect to $\mu$ at fixed unrenormalised coupling
$g$, we obtain
$$
0=\beta (g_R)\Big [{\pd\over\pd g_R}\log Z(g_R)+{1\over g_R}\Big ]+\half\e
\eqno(3.3a)
$$
Differentiating it instead at fixed $\mu$, we then have
$$
{dg\over g}=\Big [{\pd\over\pd g_R}\log Z(g_R)+{1\over g_R}\Big ]dg_R=
-{\e\over 2\beta (g_R)}dg_R
\eqno(3.3b)
$$

The pressure, being a physical quantity, is the same before and after
renormalisation:
$$
P_R(T,g_R)=P(T,g)
\eqno(3.4)
$$
Of course, $P_R$ depends also on $\mu$.
Hence (2.7) is equivalent to 
$$\eqalignno{
\beta (g_R){\pd\over\pd g_R}\hat P_R(T,g_R)&=
-\quarter\e\Big [\langle F^2(0)\rangle -\langle F^2(0)\rangle_{\F}\Big ]\cr
&=\Big [ \langle T^\mu_{\;\;\mu}>-<T^\mu_{\;\;\mu}\rangle _{\F}\Big ]&
(3.5a)\cr}
$$
where we have used (2.10).
So  (2.12a) becomes
$$
\beta (g_R){\pd\over\pd g_R}\hat P_R(T,g_R)=
-\quarter\e(D^0-(3-\e )D^3)
\eqno(3.5b)
$$

Because the left-hand side of (3.5) must be finite when $\e\to 0$, the
divergent part of $(D^0-3D^3)$ 
as a function of $g_R$
must be of order $\e^{-1}$ exactly. 
Because the derivative of the pressure in (2.12b) is a physical quantity and so
must be finite, the
divergent parts of $D^0$ and $D^3$, which are temperature-dependent, must be
equal and opposite: 
$$\eqalignno{
D^0(T,g_R)&\sim {D(T,g_R)\over\e}+d^0(T,g_R)+O(\e )\cr
D^3(T,g_R)&\sim -{D(T,g_R)\over\e}+d^3(T,g_R)+O(\e )&(3.6)\cr}
$$
Again, $D^0$ and $D^3$ depend also on $\mu$.
Hence when $\e\to 0$
$$
\hbox{Lim}_{\e\to 0}{\pd\over\pd g_R}\hat P_R(T,g_R)= 
-{1\over\tilde\beta(g_R)}D(T,g_R)=-\hbox{Lim}_{\e\to 0}{\e\over \beta (g_R)} D^0
\eqno(3.7a)
$$
or, equivalently,
$$
\tilde\beta(g_R){\pd\over\pd g_R}\hat P_R(T,g_R)=-D(T,g_R)=\langle
T^\mu_{\;\;\mu} \rangle
\eqno(3.7b)
$$
We have used the fact that $<T^{\mu}_{\mu}>_{\F}$ vanishes in the limit
$\e\to 0$.

We now use the renormalisation group to show that the thermodynamic
pressure $P$ is equal to the hydrodynamic pressure $\cal P$ 
defined in (1.4).
Introduce the running coupling $\bar g(T;g_R)$ such that 
$$
T{\pd\over\pd T}\bar g(T;g_R)\Big\arrowvert _g
=\beta (g_R){\pd\over\pd g_R}\bar g(T;g_R)\Big\arrowvert _T
~~~~~~~~~~~~~\bar g(\mu;g_R)=g_R
\eqno(3.8a)
$$
It is familiar that, to calculate $\bar g(T;g_R)$, one makes use
of the identity
$$
\beta (g_R){\pd\over\pd g_R}\bar g(T;g_R)\Big\arrowvert _T
=\beta (\bar g(T;g_R))
\eqno(3.8b)
$$
Because $\hat P_R$ has dimension $T^n$, we may write
$$
\hat P_R(T,g_R)=T^n\phi (g_R,T/\mu)
\eqno(3.9)
$$
{}From renormalisation group theory,
$$
\phi (g_R,T/\mu)=\phi (\bar g(T;g_R),1)
\eqno(3.10)
$$
Thus 
$$
T{\pd\over\pd T}\phi (g_R,T/\mu)=\beta (\bar g(T;g_R)){\pd\over\pd\bar g(T;g_R)}
\phi (\bar g(T;g_R),1)
\eqno(3.11)
$$
{}From (3.8b), we may replace $\beta (\bar g(T;g_R)){\pd/\pd\bar g(T;g_R)}$
with $\beta (g_R){\pd/\pd g_R}$. Hence
$$
T^{n+1}{\pd\over\pd T}\big [T^{-n}\hat P_R(T,g_R)\big ]
=\beta (g_R){\pd\over\pd g_R} \hat P_R(T,g_R)
\eqno(3.12)
$$
With this equation, together with (2.8) and (3.5a), we obtain
$$\eqalignno{
\hat P_R&=-{1\over n-1}\Big\{
\big [\langle T^\mu_{\;\;\mu}(0)\rangle-\langle T^\mu_{\;\;\mu}(0)\rangle_{\F}
\big ]-
\big [\langle T^{00}(0)\rangle-\langle T^{00}(0)\rangle_{\F}\big ]\Big\}\cr
&={1\over n-1}\big [\langle T^{ii}(0)\rangle
-\langle T^{ii}(0)\rangle_{\F}\big ]\cr
&=\big [\langle T^{33}(0)\rangle-\langle T^{33}(0)\rangle_{\F}\big ]&
(3.13)\cr}
$$
where in the last step we have used the spherical symmetry.
{}From (1.4), this is just the statement that the thermodynamic pressure
$P$ is equal to the hydrodynamic pressure $\cal P$.
This result is valid in spite of the presence of the trace anomaly, so
this closes a gap in the conventional proof that explicitly ignores 
it\ref{\landsman}.

This result is valid in spite of the presence of the trace anomaly.
However, the anomaly does influence the exact form of the pressure.
In terms of the quantities introduced in (3.6), 
in the limit $\e\to 0$
$$
\hat  P =-\quarter (d^0+d^3-D) 
\eqno(3.14)
$$
The last term is just 
$$
\quarter D=-\quarter\hbox{ Lim }_{\e\to0}\langle T^{\mu}_{\;\;\mu}\rangle
\eqno(3.15)
$$

We note that (3.12) and (3.5a) give also
$$
T{\pd\over\pd T}\hat P_R(T,g_R)-(4-\e )\hat P_R(T,g_R)=
\big [\langle T^{\mu}_{\mu}(0)\rangle-\langle T^{\mu}_{\mu}(0)\rangle_{\F}\big ]
\eqno(3.16)
$$
so when $\e\to 0$
the departure of the pressure from simple $T^4$ behaviour is just the
result of the QCD trace anomaly.
(The other parts of the energy-momentum tensor,
arising from the gauge fixing and the ghosts, do not contribute to this
trace anomaly\defref\collins{
J C Collins, A Duncan and D Joglekar, Physical Review D16 (1977) 438
}.)

\bigskip\goodbreak
{\bf 4 Temporal gauge}

In the gauge $A^0=0$ the expression for $D^0$ in terms of the gluon field
is particularly simple:
$$
D^0 =-\Big [\langle (\pd ^0 A^i)^2\rangle-\langle (\pd ^0 A^i)^2\rangle _{\F}
\Big]
\eqno(4.1)
$$
There are some unanswered questions about the temporal gauge, both at
zero temperature\defref\huffel{
H H\"uffel, P V Landshoff and J C Taylor, Physics Letters B217 (1989) 147
}
and more particularly in the finite-temperature imaginary-time
formalism\defref\imtime{
K A James and P V Landshoff, Physics Letters B251 (1990) 167
},
though the real-time formalism may well be free of problems\defref\james{
K A James, Z Physik C48 (1990) 169
}. 
In the temporal gauge, we can ignore Faddeev-Popov ghosts. 

With the Keldysh contour,
$$
D^0(0)=\int {d^{4-\e}q\over (2\pi )^{4-\e}}(q^0)^2[
\D^{ii}_{12}(q)-\D^{ii}_{12\F}(q)]
\eqno(4.2)
$$
where the subscript {\sevenrm 12} refers to the element of the $2\times 2$
thermal matrix propagator\ref{\lebellac}. This matrix has the structure
$$
{\bf \D}^{ij}(q)={\bf M}(q^0)\Big \{\Big (\delta ^{ij}-{q^iq^j\over {\bf q}^2}
\Big )\tilde{\bf \D}_T(q) 
+{q^iq^j\over {\bf q}^2}\tilde{\bf \D}_L(q)\Big\}  {\bf M}(q^0)
\eqno(4.3)
$$
where the matrices $\tilde{\bf \D}_T$ and $\tilde{\bf \D}_L$ are diagonal
and
$$
{\bf M}(q^0) =  \sqrt{n(q^0)} \left[\matrix{ e^{{1\over 2} \beta |q^0|} &
e^{-{1\over 2}\beta q^0} \cr e^{{1\over 2}\beta q^0} & e^{{1\over
2}\beta |q^0|}\cr}\right ]
\eqno(4.4)
$$
with $n(q^0)$ the Bose distribution $(e^{\beta |q^0|}-1)^{-1}$.
To zeroth order in the coupling
$$\eqalignno{
\tilde{\bf \D}_T(q) = &  \left[ \matrix{\D_T(q)&0\cr
                   0&  \D_T^*(q)\cr}\right ]
~~~~~~~~~~~~~~~~~~~\D_T(q)={i\over {q^2 + i\eta}} \cr
&~~\cr
\tilde{\bf \D}_L(q) = &  \left[ \matrix{\D_L(q)&0\cr
                   0&  \D_L^*(q)\cr}\right ]
~~~~~~~~~~~~~~~~~~~\D_L(q)={i\over (q^0)^2} &(4.5)\cr}
$$
where some prescription is needed\ref{\huffel}
to define the meaning of $1/(q^0)^2$.
As long as a rotationally invariant pole prescription is used,
the Dyson resummed propagator has the same structure as in (4.3), but with 
$$\eqalignno{
&\D_T(q)={i\over q^2 -\Pi _T(q,\beta)} 
\cr
&\D_L(q)={i\over (q^0)^2-\Pi _L(q,\beta)} 
 &(4.6)\cr}
$$
We need not retain the appropriate prescriptions $\Delta_L$ in (4.5);
they play no part because 
the self energies $\Pi _T$ and $\Pi _L$ have zero imaginary parts 
only at $q^0=0$, and when we use (4.6) in (4.2) the contribution from
this point is killed by the $(q^0)^2$ in the numerator of the integrand.
When we insert (4.2) into (3.7) we obtain
$$\eqalignno{
\beta(g_R)&{\pd\over\pd g_R}\hat P_R(T,g_R)\cr
&=8\e\hbox{ Im }\int {d^{4-\e}q\over (2\pi )^{4-\e}}\theta (q^0)(q^0)^2
\big (1+2n(q^0)\big )\Big\{{2\over q^2 -\Pi _T(q,\beta)}-{2\over q^2}
+{1\over (q^0)^2-\Pi _L(q,\beta)}-{1\over (q^0)^2}\Big\}\cr
&&(4.7)\cr}
$$
The factor of 8 is included because a
trace over colour indices was understood in all the equations.
As in our previous work\ref{\foam}, any infrared
divergence problem has been rendered harmless by the Dyson resummation:
divergences of the self-energies are irrelevant because they now appear in
denominators, and any zero of the denominators at $q=0$ is unimportant
because  of the powers of $q$ present in $d^{4-\e}q$.

In making use of (4.7) in a semi- or non-perturbative way, 
which is the goal of subsequent work,
we have to renormalise accordingly. 
It is known \defref\renorm{
W Kummer, Acta Physica Austriaca 41 (1975) 315
}
that removing a factor $Z^{-2}$ from the first and third terms of the
integral (4.7) renders their contribution to the 
integrand finite. This factor may be found by solving the differential
equation (3.3a), with the boundary condition that $Z(0)=1$. The solution
may be written in the form
$$
\log Z(g_R)+\half\log\Big ({2\beta (g_R)\over g_R\e}\Big )=
2\int _0^{g_R}d\g{\tilde\beta (\g)/\g^2-
\half(\pd/\pd \g)(\tilde\beta (\g)/\g)\over
  \e-2\tilde\beta (\g)/\g}
\eqno(4.8)
$$
This integral converges when $\e\to 0$, so the divergent part of $Z^{-2}$
is
$$
{1\over Z^2}\sim -{2\beta (g_R)\over g_R\e}
\eqno(4.9)
$$
So (4.7) finally becomes
$$\eqalignno{
{\pd\over\pd g_R}\hat P_R(T,g_R)
=-16\hbox{ Lim }_{\e\to0}\hbox{ Im }&\int 
{d^{4-\e}q\over (2\pi )^{4-\e}}\theta (q^0)(q^0)^2
\big (1+2n(q^0)\big )\cr
&\Big\{{2\over q^2 -\Pi ^C _T(q,\beta)}-{2\over q^2}
+{1\over (q^0)^2-\Pi ^C _L(q,\beta)}-{1\over (q^0)^2}\Big\}
&(4.10)\cr}
$$
where $\Pi ^C_T$ and $\Pi ^C_L$ are convergent thermal self-energies.

\bigskip
{\bf 5 Small-coupling limit}

So far, our formulae are exact: we have not used perturbation theory to
derive them. We now investigate how they behave in the limit of small
coupling. We content ourselves with simple results that follow
without detailed calculation, and leave a more extensive investigation
to a future paper.

Because\ref{\braaten}\defref\shuryak{B A Freedman and L D McLerran,
Physical Review D16 (1977) 1130, 1147, 1169\hfill\break
E Shuryak, JETP 47 (1978) 212\hfill\break
J I Kapusta, Nuclear Physics B148 (1979) 461
}
$\hat P_R(T,g_R)$ behaves as $-\textstyle{{1\over 6}}g_R^2T^4$
for small $g_R$, we see from (3.2b)  and (3.7b) that
$$
D=O(g_R^4)
\eqno(5.1)
$$
We may obtain almost the same information directly from (2.12a), which
becomes with (3.6) 
$$
{\pd\over\pd g}\hat P_R(T,g)={1\over 2g}\Big [{4D\over\e}+d^0-3d^3-D
+O(\e )\Big ]
\eqno(5.2)
$$
To lowest order the unrenormalised and renormalised couplings are equal,
so to this order the left-hand side is again ${\pd /\pd g_R}\hat P_R(T,g_R)$;
therefore the singular term on the right-hand side must be at least of
order $g^3_R$. Furthermore, in order to give agreement with (3.7b),
$$
d^0-3d^3=-{2g_RD\over\tilde\beta(g_R)}+\dots
\eqno(5.3)
$$
where the further terms are at least of order $g_R^3$.

The pressure may also be calculated directly from (3.14).
For this to be compatible with (5.2) in lowest order, it must be that
$d^0=d^3$ to lowest order. That is
$$
\hat P_R=-\half d^0+\dots
\eqno(5.4)
$$
where the further terms are of order $g_R^3$ at least.

\bigskip
{\bf 6 Conclusion}

(3.7) is an exact relation between the QCD thermal pressure and the thermal
average of the trace anomaly.  We have used this to prove nonperturbatively
the equivalence of the thermodynamic and the hydrodynamic pressure.

With the plausible assumption that thermal field theory may be
formulated consistently in the temporal gauge, (3.7) can be expressed in 
terms of a manifestly
infrared finite integral (4.10) over the full propagator. 
The absence of a chromomagnetic screening mass 
poses no particular problem to an evaluation of (4.10), so it
may provide a suitable starting point for developing new resummation
techniques.

\bigskip
\bigskip
{\eightit
This research is supported in part by the EU Programme
``Training and Mobility of Researchers", Networks
``Hadronic Physics with High Energy Electromagnetic Probes"
(contract FMRX-CT96-0008) and
``Quantum Chromodynamics and the Deep Structure of
Elementary Particles'' (contract FMRX-CT98-0194), and by PPARC.
One of us (PVL) is grateful also  to the International Erwin-Schr\"odinger
Institute for its hospitality.}
\vfill\eject
\medskip\immediate\closeout\rfile\writestoppt
\baselineskip=14pt{{\bf References}}\bigskip{\frenchspacing%
\parindent=20pt\escapechar=` \input refs.tmp\bigskip}\nonfrenchspacing
\bye